\def \SAIT #1 #2 {{\em Mem.\ Soc.\ Astron.\ It.\/} {\bf #1}, #2}
\def \MESS #1 #2 {{\em The Messenger\/} {\bf #1}, #2}
\def \ASTRNACH #1 #2 {{\em Astron. Nach.\/} {\bf #1}, #2}
\def \AAP #1 #2 {{\em Astron. Astrophys.\/} {\bf #1}, #2}
\def \AAL #1 #2 {{\em Astron. Astrophys. Lett.\/} {\bf #1}, L#2}
\def \AAR #1 #2 {{\em Astron. Astrophys. Rev.\/} {\bf #1}, #2}
\def \AAS #1 #2 {{\em Astron. Astrophys. Suppl. Ser.\/} {\bf #1}, #2}
\def \AJ #1 #2 {{\em Astron. J.\/} {\bf #1}, #2}
\def \ANNREV #1 #2 {{\em Ann. Rev. Astron. Astrophys.\/} {\bf #1}, #2}
\def \APJ #1 #2 {{\em Astrophys. J.\/} {\bf #1}, #2}
\def \APJL #1 #2 {{\em Astrophys. J. Lett.\/} {\bf #1}, L#2}
\def \APJS #1 #2 {{\em Astrophys. J. Suppl.\/} {\bf #1}, #2}
\def \APSS #1 #2 {{\em Astrophys. Space Sci.\/} {\bf #1}, #2}
\def \ASR #1 #2 {{\em Adv. Space Res.\/} {\bf #1}, #2}
\def \BAIC #1 #2 {{\em Bull. Astron. Inst. Czechosl.\/} {\bf #1}, #2}
\def \JSQRT #1 #2 {{\em J. Quant. Spectrosc. Radiat. Transfer\/} {\bf #1}, #2}
\def \MN #1 #2 {{\em Mon. Not. R. Astr. Soc.\/} {\bf #1}, #2}
\def \MEM #1 #2 {{\em Mem. R. Astr. Soc.\/} {\bf #1}, #2}
\def \PLR #1 #2 {{\em Phys. Lett. Rev.\/} {\bf #1}, #2}
\def \PASJ #1 #2 {{\em Publ. Astron. Soc. Japan\/} {\bf #1}, #2}
\def \PASP #1 #2 {{\em Publ. Astr. Soc. Pacific\/} {\bf #1}, #2}
\def \NAT #1 #2 {{\em Nature\/} {\bf #1}, #2}

\def\sp{\hspace{1.5pt}}
\def\etal{{\it et~al.}}
\def\amin{\ifmmode^{\prime}\else$^{\prime}$\fi}
\def\asec{\ifmmode^{\prime\prime}\else$^{\prime\prime}$\fi}

\def\simgt{\lower.5ex\hbox{$\; \buildrel > \over \sim \;$}}
\def\simlt{\lower.5ex\hbox{$\; \buildrel < \over \sim \;$}}

\newcommand\rcw{\hbox{RCW\hspace{1.5pt}103}}
\newcommand\src{\hbox{1E\hspace{1.5pt}161348$-$5055}}

\newcommand\ASCA{{\it ASCA}}
\newcommand\asca{{\it ASCA}}

\newcommand\rosat{{\it ROSAT}}
\def\sp{\hskip 1.5pt}

\def\psr{\hbox{1E 1841$-$045}}
\def\newpsr{\hbox{PSR J1845$-$0258}}

\documentstyle{memsait}
\input epsf.sty
\begin{opening}
\title{RADIO-QUIET X-RAY PULSARS IN SUPERNOVA REMNANTS AND THE ``MISSING'' PULSAR PROBLEM} 

\author{E. V. Gotthelf}
\institute{NASA/Goddard Space Flight Center, Code 662, Greenbelt, MD 20771}

\date{} 
\end{opening}

\begin{document}

\oddpagefooter{}{}{} 
\evenpagefooter{}{}{} 

\bigskip

\begin{abstract}

The paradigm that young neutron stars (NSs) evolve as rapidly
rotating Crab-like pulsars requires re-examination. Evidence is
accumulating that, in fact, many young NS are slowly rotating ($P \sim
10$-s) X-ray pulsars, lacking in detectable radio emission. We present
new results on three radio-quiet NS candidates associated with
supernova remnants, which suggests that alternative
evolutionary-paths exist for young pulsars. These include the 12-s
pulsator in Kes 73, the 7-s pulsar near Kes 75, and the enigmatic
X-ray source in RCW 103.  We postulate that such objects account for
the apparent paucity of radio pulsars in supernova remnants.

\end{abstract}

\section{Where Are All The Young Neutron Stars?}

Neutron stars are thought to be born as rapidly rotating ($\sim 10$
ms) radio pulsars created during a Type II/Ib supernova explosion
involving a massive star. Their existence was postulated in 1934 by
Baade \& Zwicky (1934) based on theoretical arguments, but had to wait
until the 1970s for observational support, provided by the remarkable
discoveries of the Crab and Vela pulsars in their respective supernova
remnants (SNRs).

The properties of these pulsars were found to be uniquely explained in
the context of rapidly rotating, magnetized neutron stars emitting
beamed non-thermal radiation. Their fast rotation rates and large
magnetic fields ($\sim 10^{12}$ G) are consistent with those of a
main-sequence star collapsed to NS dimension and density. A fast
period essentially precluded all but a NS hypothesis and thus provided
direct evidence for the reality of NSs (see Shapiro \& Teukolsky 1983 for
a brief history and intro to NS physics). Furthermore, their
inferred age and association with SNRs provided strong evidence that
NSs are indeed born in supernova explosions.  These properties were
considered typical of all young pulsars, but as we shall see, there
is new evidence that suggests that this is unlikely to be the case.

Most supernovae (non Type Ia) are expected to produce a NS, whose 
unpulsed emission
should be easily discernible in the radio-band during the lifetime of
a typical SNR ($\simgt 10^{4}$ yrs) as a radio-loud ``plerion''
(Weiler
\& Sramek 1988). So it is quite remarkable that, despite detailed 
radio searches, few of the hundreds of known SNRs have yielded a NS
candidate. Furthermore, comprehensive radio surveys suggest that most
radio pulsars near SNRs can be attributed to chance overlap
(e.g. Lorimer \etal\ 1998; Gaensler \& Johnston 1995; see
Kaspi \etal\ 1996 for a review).  With the results of these new
surveys, traditional arguments for the lack of observed radio pulsars
associated with SNR, such as those invoking beaming and large ``kick''
velocities, are less compelling.

It is now clear that this discrepancy is an important and vexing
problem in current astrophysics.

\section{The Revolution Evolution: Slowly Rotating Young X-ray Pulsars}

Progress in resolving this mystery is suggested by X-ray observations
of young SNRs. These are revealing X-ray bright, but radio-quiet
compact objects at their centers. It is now understood that these
objects form a distinct class of radio-quiet neutron stars (Caraveo et
al. 1996, Gotthelf, Petre, \& Hwang 1997; Vasisht et al. 1997; and
refs. therein), perhaps born or evolving in a fashion drastically
different from than of the Crab.

Some of these sources have been found to be slowly rotating pulsars
with unique properties. Their temporal signal is characterized by spin
periods in the range of $5 - 12$ s, steady spin-down rates, and highly
modulated sinusoidal pulse profiles ($\sim 30\%$). They have steep
X-ray spectra (photon index $\simgt 3$) with X-ray luminosities of
$\sim 10^{35}$ erg cm$^{-2}$~s$^{-1}$. As a class, these seemingly
isolated pulsars are currently referred to as the anomalous X-ray
pulsars (AXP; van Paradijs et al. 1995). Nearly half are located at
the centers of SNRs, suggesting that they are relatively young ($
\simlt 10^{5}$ yr-old). And so far, no counterparts at other 
wavelengths have been identified for these X-ray bright objects.  The
prototype for this class, the 7-s pulsar 1E 2259+586 in the $\sim
10^4$ yr old SNR CTB\sp109, has been known for nearly two decades
(Gregory \& Fahlman 1980).

These are now about a dozen slow X-ray pulsars apparently associated
with young SNRs (originally Gregory \& Fahlman 1980; see also Table I,
Gotthelf \& Vasisht 1998 for a recent summary). These include the four
known soft $\gamma$-ray repeaters (SGR), also likely to be associated
with young SNRs (Cline et al. 1982; Kulkarni \& Frail 1993; Vasisht et
al. 1994), which have recently been confirmed as slow rotators
(Kouveliotou \etal\ 1998).  {\bf  In fact, there are currently more
known slow, radio-quiet X-ray pulsars in the center of identified SNR
than confirmed Crab-like radio pulsars!}

Here, we present new results on three intriguing radio-quiet, X-ray
bright neutron star candidates which we are studying closely.  These
include the recently discovered 12-s X-ray pulsar in the SNR Kes\sp73,
very likely to be an isolated ``magnetar'', a pulsar with an enormous
magnetic field (B $\sim 10^{14}$ G); the newly discovered (March
1998), bright 7-s pulsar near Kes 75, AX~J1845-03, which displays
similar properties; and a follow-up \asca\ observation of RCW~103,
which helps resolve some long standing mysteries about this enigmatic
object. The study of these and closely related objects are shedding
new light on the evolution of young NSs.

\subsection{The Remarkable X-ray Pulsar in Kes\sp73}

The recent discovery of pulsed X-ray emission from the central compact
source in SNR Kes\sp73 (Vasisht \& Gotthelf 1997) was somewhat
surprising, as this unresolved Einstein source, 1E\sp1841$-$045, has
been studied for some time (Kriss \etal\ 1985; Helfand \etal\
1994). The slow (12-s) period is most unusual for a young pulsar; if
this is an ``isolated'' neutron star, then it is the one having the
longest spin period ever observed.

This pulsar was initially detected during a 1993 \asca\ X-ray
observation of Kes\sp73 and confirmed with a weak archival \rosat\
detection, which indicated a unusually rapid spin-down rate (Vasisht \&
Gotthelf 1997). We have recently obtained a new \asca\ measurement of
the pulsator which provides irrefutable confirmation of its remarkable
spin-down. 

Fig. 1 shows the periodigram for the two \asca\ epochs (From Gotthelf
\& Vasisht 1998, in prep.).  The pulsar is apparently spinning down
rapidly at a rate of $4.1 \times 10^{-11}$ s/s, orders of magnitude
faster than the Crab-like pulsars. This rate is consistent with that
found using the \rosat\ data and suggests a linear trend.  Most
importantly, the characteristic pulsar age is consistent with the age
derived for the SNR ($\sim 2,000$ yrs). The flux has remained steady
between the two ASCA observations, consistent with the \rosat\ flux
measurement. And the spectrum also remains unchanged, described by a
steep power-law of photon index $\sim 3.4$, unlike a typical Crab-like
or recycled pulsar.

These properties have important consequences for a young pulsar.  The
rotational energy of 1E\sp1841$-$045 is far too small to power its
total X-ray emission of $L_X \sim 3\times10^{35}d_7^2$. The maximum
luminosity derivable just from spin-down is $L_X \simlt 4\pi^2I \dot
P/ P^3 \sim 10^{34}$ erg/s. On the other hand, the observed
luminosity is appreciably low for an accretion powered binary system
$\sim 10^{36-38}$ erg/s. However, the derived luminosity, steep
spectrum, and lack of stochastic variability makes an accretion scenario
all but unlikely. In \S 3 we discuss the several models to explain the
nature of this and other AXPs.

\bigskip
\bigskip
\begin{figure}[here]
\epsfxsize=6.0cm
\centerline{\hfill \epsfbox[50 50 500 500]{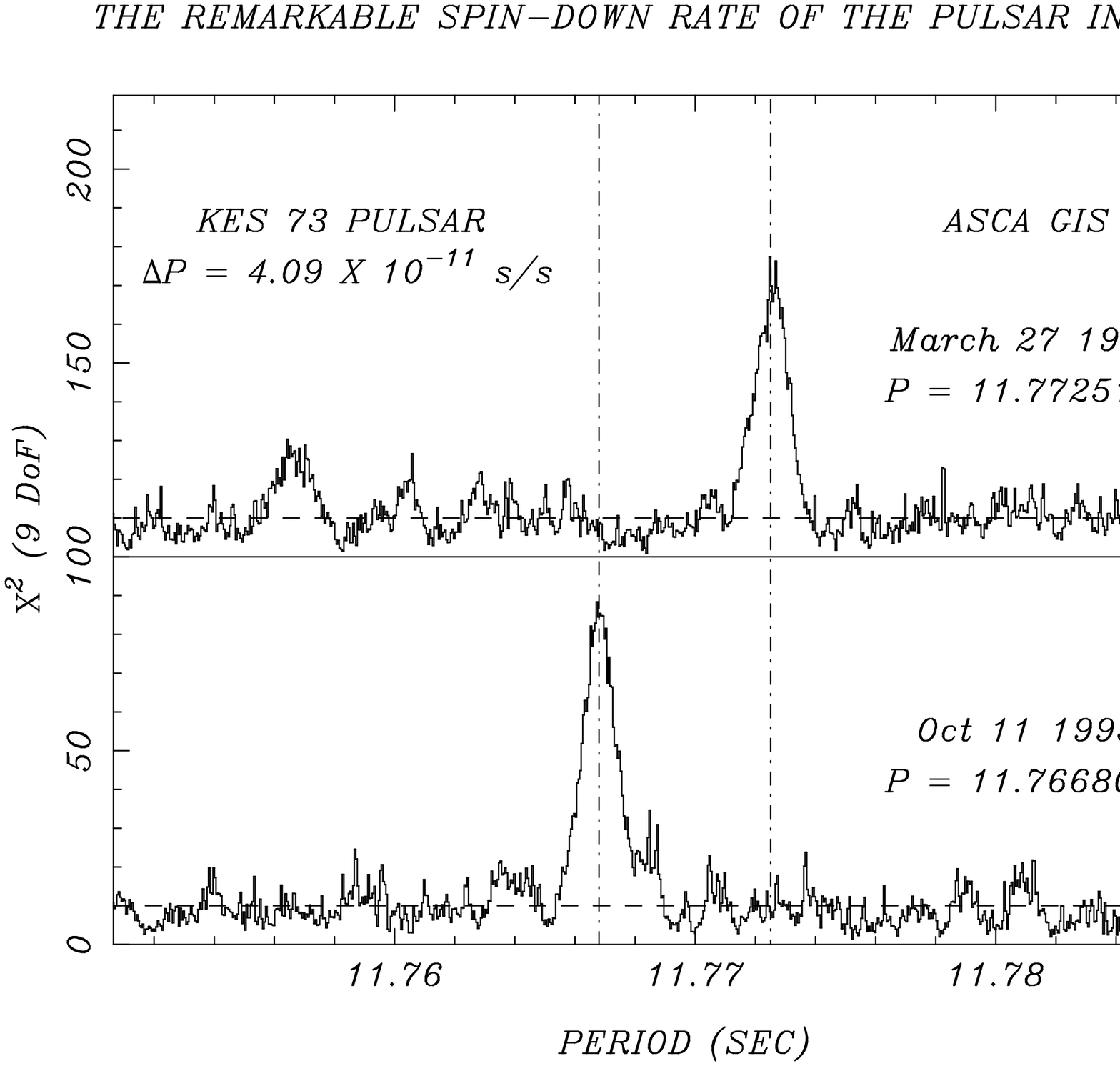}  \hfill}
\caption{The \asca\ spin-down history of the Kes~73 pulsar. (Lower panel) The Oct 11 1993 GIS detection of the 12 sec pulsar in Kes~73. (Upper
panel) The follow-up March 27 1998 GO observation. The period
derivative, assuming a linear trend, is $4.1 \times 10^{-11}$ s/s.}
\end{figure}

\subsection{\newpsr: A 7 Second Anomalous Pulsar in the Distant Milky Way}

A recent automated search through the \ASCA\ archive (Gotthelf \&
Vasisht 1998; Torii 1998) has revealed another AXP-like slow
pulsator. Like the 12 s pulsar in Kes\sp73, this 7 s rotator has a
sinusoidal pulse shape and is modulated with a 35\% amplitude. The
pulsar lies at the edge of the ASCA field containing the SNR Kes
75, but their separation makes an association most unlikely.

\begin{figure}[here]
\epsfxsize=6.0cm
\centerline{\hfill \epsfbox[50 50 500 500]{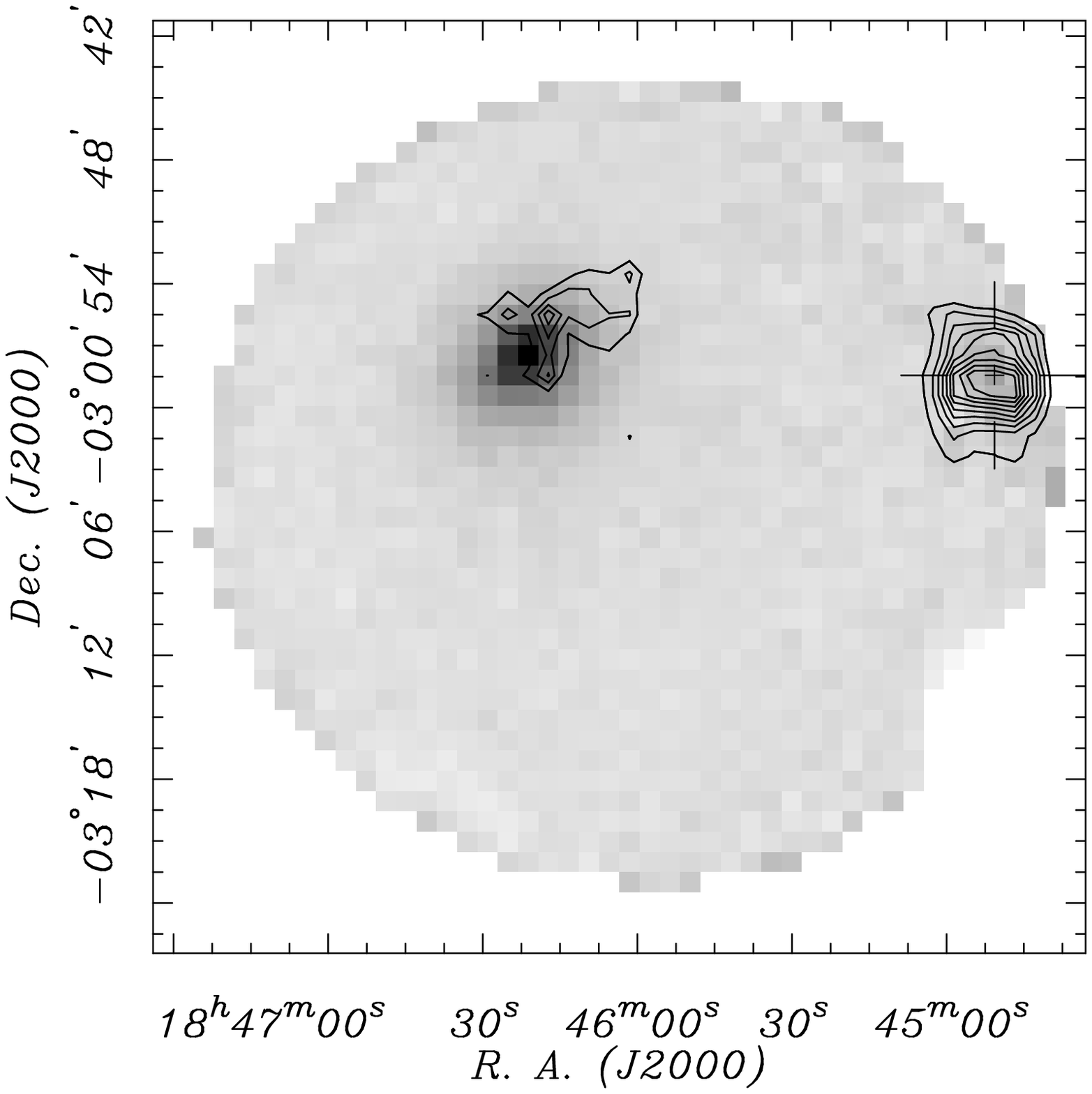} \hspace{1cm} \epsfxsize=5.0cm \epsfbox[175 500 400 800]{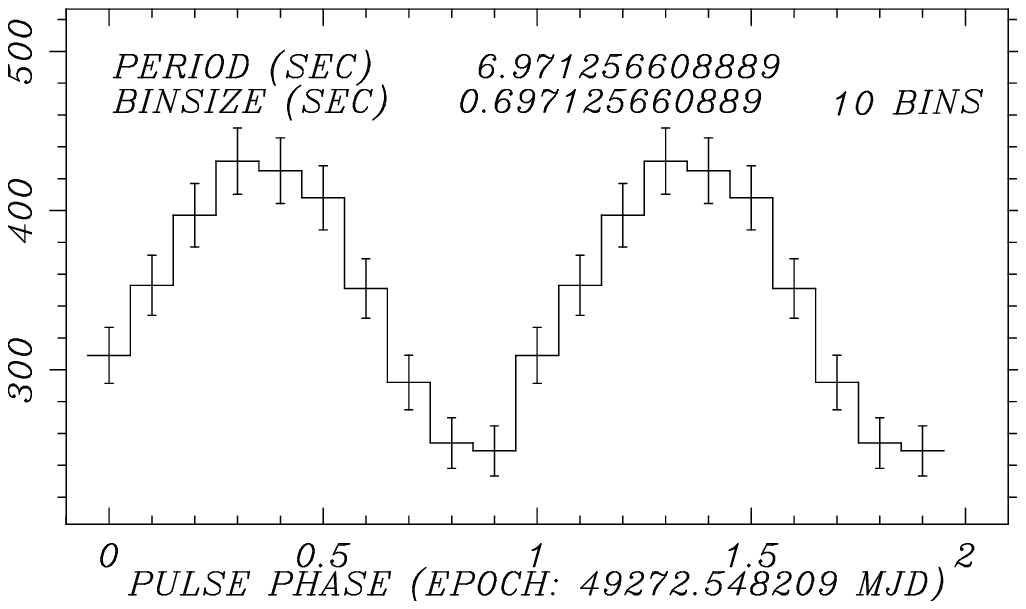} \hfill}

\caption{Discovery of a 7 s X-ray pulsar near Kes 75. (Left) The GIS pulsed 
emission image (contours) overlayed on the phase integrated image (greyscale). 
Kes 75 is on the left and \newpsr\ on the right. (Right) The 
folded GIS pulse profile of \newpsr. Two cycles are shown for clarity.  }
\end{figure}

On the basis of its properties, \newpsr\ is most likely the latest
example of an anomalous X-ray pulsar. The period is nearly identical
to that of the pulsar in \hbox{CTB\sp109}. The light curve is consistent with
that of a steady source. A preliminary search for variability fails to show
Doppler modulation which would signal the presence of a binary companion. In addition,
the low frequency spectrum shows no obvious pink noise typical
of accretion powered sources. The X-ray spectrum is remarkably
steep, with a power law index $\sim 5$. The steep index is indicative of the
tail of an intrinsically thermal spectrum. An absorbed blackbody also
gives a good fit with a temperature of $kT \sim 0.64$ keV, similar
to that found for other AXPs. 

A high foreground absorption suggests that the pulsar is distant and
likely highly absorbed.  Uncertainty in the distance to the pulsar
makes the luminosity difficult to estimate. Our best guess is 15 kpc, an estimate likely to be accurate to within a
factor of two. The isotropic X-ray luminosity is then $L_X \sim
2 \times 10^{35}d_{15}^2$ erg s$^{-1}$. Note that like for the Kes\sp73
pulsar, the spectrum of \newpsr\ is unlike that of any accreting
high-mass neutron star binary. 

We await further multi-wavelength observations of \newpsr\ necessary
to fully describe this source and gauge its importance. There is
evidence in the MOST radio survey data for an underlaying shell-type
SNR (B. M. Gaensler, Priv Comm.).  If verified, this would be a
spectacular endorsement of the AXP hypothesis.  Although only a future
$\dot P$ measurement can help determine its spin-down age.

\subsection{The Neutron Star Candidate \src}

The central X-ray source in \rcw, \src, has also been known for some
time (Tuohy \& Garmire 1980), but unlike Kes\sp73, so far no pulsar
has been identified.  Nevertheless, several other properties of this
source identify it as AXP-like object. Located in the center of a
young SNR, the compact source has an unexpectedly steep spectral
signature (photon index $\sim 3$) with an X-ray luminosity of $\sim
10^{35}$ erg cm$^{-2}$ s$^{-1}$ (Gotthelf, Petre \& Hwang 1997). But
the lack of detected pulsations leave the nature of \src\ ambiguous
and so far no clear interpretation of the origin of the X-ray emission
has emerged.

Progress on this front is suggested by a recent follow-up \asca\
observation of \rcw. The flux from \src\ was seen to decrease by a
factor of $\sim 10$ compared to the 1993 observation 3.5 yrs earlier
(see Fig. 1, Petre \& Gotthelf 1998, this Proceedings). This was quite
unexpected as the count rates for the HRI detections at two epochs
spanning 16 yrs agreed to within $\sim 10\%$.  The spectral fits to
the original \asca\ data only allowed a flux, at most, $<3$ times higher
that predicted by the HRI measurements.

Our new result, the detection of significant variability from \src,
strongly rules out a simple cooling NS origin, as originally postulated
by Tuohy \etal\ (1993). A fuller account of this result is
presented elsewhere in this Proceedings (Petre \& Gotthelf 1998).
In the next section, we present several models which may help to
elucidate the aforementioned objects, and discuss their significance.
 
\section{Interpretation and Discussion}

The spectral and temporal properties of \psr\ and \newpsr\ are similar
to each other and to the other seemingly isolated, young AXPs. And
these objects appear related to the SGRs. Similarly, although no pulsation
have been found from central source in \rcw, its spectral properties
suggest a relation to these slow pulsars.

If these and other NS candidates like them were indeed born as 
fast rotators, then a mechanism must be found to slow them down to
their currently observed rates. The rapid but steady spin-down 
of the Kes\sp73 pulsar suggests a possibility. The
equivalent magnetic field for a rotating dipole is $B_{dipole}
\simeq 3.2\times10^{19}~(P\dot P)^{1/2} \approx 8\times 10^{14}$ G,
one of the highest magnetic fields observed in nature.  Theory
describing a NS with such an enormous field, a ``magnetar'', has been
worked out by Duncan \& Thompson (1992). Vasisht \& Gotthelf (1997)
suggest that the Kes\sp73 pulsar was born as a magnetar $\sim
2\times10^3$ yrs ago and has since spun down to a long period due to
rapid dipole radiation losses.  This would then be the first direct
evidence of such a magnetar (see Vasisht \& Gotthelf 1997 for a detail
discussion). In the context of the above theory, \src\ may have been
born a magnetar which has subsequently spun down.

Alternatively, some NSs may be born as slow ($\sim$ 2 s) rotators as
considered by Spruit \& Phinney (1998), resulting from strong physical
coupling between the progenitor's slowly rotating envelope and its
pre-collapse core.

The enormous magnetic field postulated for these slow pulsars may
provide a natural explanation for their radio-quiet nature. It has been
argued that for supercritical fields, magnetic pair creation
$\gamma\to e^+e^-$, the source of electrons for radio emission, is
suppressed (Baring \& Harding 1998).

In now seems likely that the Crab and Vela pulsars are in fact rare, 
but highly visible, examples of young NS evolution. They might
have obtained their rapid initial spins through some impulsive
mechanism that imparts both kicks and spins to young neutron stars
(Spruit \& Phinney 1998). Slow AXP pulsars, exemplified by Kes\sp73,
are likely to be more common, but observationally less obvious
manifestations of young NSs.  The SGRs may represent an
evolutionary stage during which young NSs are likely to be produce
bursts. Under this scenario, the AXPs and SGR phenomena
are closely related, linked by their strong magnetic field.

The discovery of new examples of slow pulsars is revolutionizing our
understanding young NS evolution -- it suggests that an alternative
evolution scenario for young NSs, is not only possible, but most
probable. We consider that many of the young NSs ``missing'' in
radio surveys can be accounted for by the above discussed radio-quiet
NSs.

\acknowledgements
We thank B. M. Gaensler for bringing to our attention the possibility
of a faint MOST radio shell-type SNR at the location of \newpsr.


\end{document}